\documentclass[twocolumn,aps,prl,10pt,amsmath,amssymb,nofootinbib,showpacs,superscriptaddress,floatfix]{revtex4-1}

\newcommand{\rs}{\rm\scriptscriptstyle}

\usepackage{graphicx,epsfig}
\usepackage{amsmath,amssymb,bbm}
\usepackage{color}
\usepackage[usenames,dvipsnames]{xcolor}
\usepackage[colorlinks=true,linkcolor=Red,citecolor=Green,linktoc=page]{hyperref}
\usepackage{multirow}

\DeclareFontFamily{U}{rcjhbltx}{}
\DeclareFontShape{U}{rcjhbltx}{m}{n}{<->rcjhbltx}{}
\DeclareSymbolFont{hebrewletters}{U}{rcjhbltx}{m}{n}

\usepackage{graphicx}
\usepackage{color}
\usepackage[usenames,dvipsnames]{xcolor}
\usepackage[colorlinks=true,linkcolor=Red,citecolor=Green,linktoc=page]{hyperref}
\usepackage{multirow}
\usepackage{float}
\usepackage{flushend}
\usepackage{balance}
\usepackage[varg]{txfonts}
\usepackage{ulem}
\usepackage{fancyhdr}

\DeclareMathSymbol{\lamed}{\mathord}{hebrewletters}{108}

\begin{document}
\title{Effective magnetic monopole mechanism for localized electron pairing in HTS
}


\author{M.\,C.\,Diamantini}
\affiliation{NiPS Laboratory, INFN and Dipartimento di Fisica e Geologia, University of Perugia, via A. Pascoli, I-06100 Perugia, Italy}

\author{C.\,A.\,Trugenberger}
\affiliation{SwissScientific Technologies SA, rue du Rhone 59, CH-1204 Geneva, Switzerland}

\author{V.\,M.\,Vinokur\,$^{\ast}$}
\affiliation{Terra Quantum AG, Kornhausstrasse 25, CH-9000 St. Gallen, Switzerland\\
	$^*$correspondence to be sent to vv@terraquantum.swiss}
\vspace{-0.9cm}
	
	\begin{abstract}
	The mechanism responsible for spatially localized strong coupling electron pairing characteristic of high-temperature superconductors (HTS) remains elusive and is a subject of hot debate. Here we propose a new 
	HTS pairing mechanism which is the binding of two electrons residing in adjacent conducting planes of layered HTS materials by effective magnetic monopoles forming between these planes. The pairs localized near the monopoles form  real-space seeds for superconducting droplets and strong coupling is due to the topological Dirac quantization condition. The pairing occurs well above the superconducting transition temperature $T_{\mathrm c}$. Localized electron pairing around effective monopoles promotes, upon cooling, the formation of superconducting droplets connected by Josephson links. Global superconductivity arises when strongly coupled granules form an infinite cluster, and global superconducting phase coherence sets in. The resulting $T_{\mathrm c}$ is  estimated to fall in the range from hundred to thousand Kelvins. Our findings pave the way for tailoring materials with elevated superconducting transition temperatures.
\end{abstract}
%
\maketitle

\vspace{-0.7cm}




\section{1. Introduction}
\noindent 
The discovery of HTS about 40 years ago promised a new era of industrial and technological advances, ranging from the power-loss free electric grids spanning continents to superconducting kitchen appliances in every household. However, materials superconducting at room temperature are still a dream. A dearth in understanding of the HTS pairing mechanism \,\cite{Barisic2013,Zaanen2015,Bozovic2020} 
hinders the desired technological progress. In HTSs, pairing has to fundamentally differ from that in conventional superconductors: the electron pairs should be spatially localized and have a size comparable or less than the distance between them\,\cite{HTS-Rev_2020}.
\smallskip

Anderson's resonating valence bond (RVB) theory of HTS based on electron correlations\,\cite{Anderson1987}, proposed recently in\,\cite{Katsnelson2016}, as well as the alternative route, deriving HTS from Cooper pairing near a quantum critical point (QCP) associated with an antiferromagnetic order-itinerant electron spin transition\,\cite{Balatsky1991,Chubukov2005}, beautifully capture many important features of superconducting and related pseudogap phases of HTS,\,\cite{Barisic2013,Zaanen2015,Matsuda2017,Bozovic2019,Taillefer2019,Mukho2019,Kivelson1998,critical,Kivelson2014,Chubukov2015,Hawthorn2016,Kivelson2019,Chubukov2020,Chubukov2020-2,Chubukov2020-3}. However, neither of them succeeded to reveal the HTS pairing mechanism.  Early attempts to induce local pairing in the Hubbard model framework \cite{add1, add2, add3, add4, add5} had to introduce an attractive interaction by hand. This, however, is different from a dynamically generated attractive potential as in BCS theory or in our model. Many aspects of the HTS, especially of the Fe-based ones, seem to be fairly well described by multi-band Hubbard models\,\cite{WangPairing,ChubukovPairing}. In this type of models, the attractive pairing interaction is genuinely different from any other one mediated by an exchange interaction and arises from the resonant hopping between the different bands \cite{HuangPairing}.  Yet, again, despite its remarkable successes\,\cite{Rybicki, Kowalski}, also for the cuprates, in reproducing the correct HTS phase diagram, even this model fails to capture essential fundamental aspects of HTS. Further, building on the relevance of quantum criticality for HTS\,\cite{Chubukov2005}, one can derive the superconducting transition temperature as $T_{\mathrm c}\sim E_{\rs F}/\sqrt{g}$, with $E_{\rs F}$ being the Fermi energy and $g$ being the coupling constant, in remarkable contrast to the standard BCS behavior, $T_{\mathrm c}\sim\exp(-{\rm const}/g)$.  Yet, a strong coupling still has to be assumed, and the origin of the characteristic feature of HTS, the small superconducting coherence length $\xi$, resulting in non-overlapping Cooper pairs, i.e., the fact that the size of the pairs is less than the inter-pairs distance\,\cite{HTS-Rev_2020}, which is opposite to what occurs in conventional superconductors is not explained. The other unexplained fundamental HTS feature is that unlike in conventional superconductors, the transition temperature of thin cuprate films does not essentially depend on film thickness and remains the same as in the bulk, see, for example\,\cite{Zhao2019}. Even an extremely thin monolayer BSCCO film containing a half of the elemental cell (but still comprising two CuO planes) has practically the same bulk $T_{\mathrm c}$\,\cite{Twoplanes}. This posits a quest for a mechanism of localized electron pairing in real space which, at the same time, remains effective even in the thinnest films of the HTS material. 
\smallskip

\begin{figure}[t!]
	\includegraphics[width=9cm]{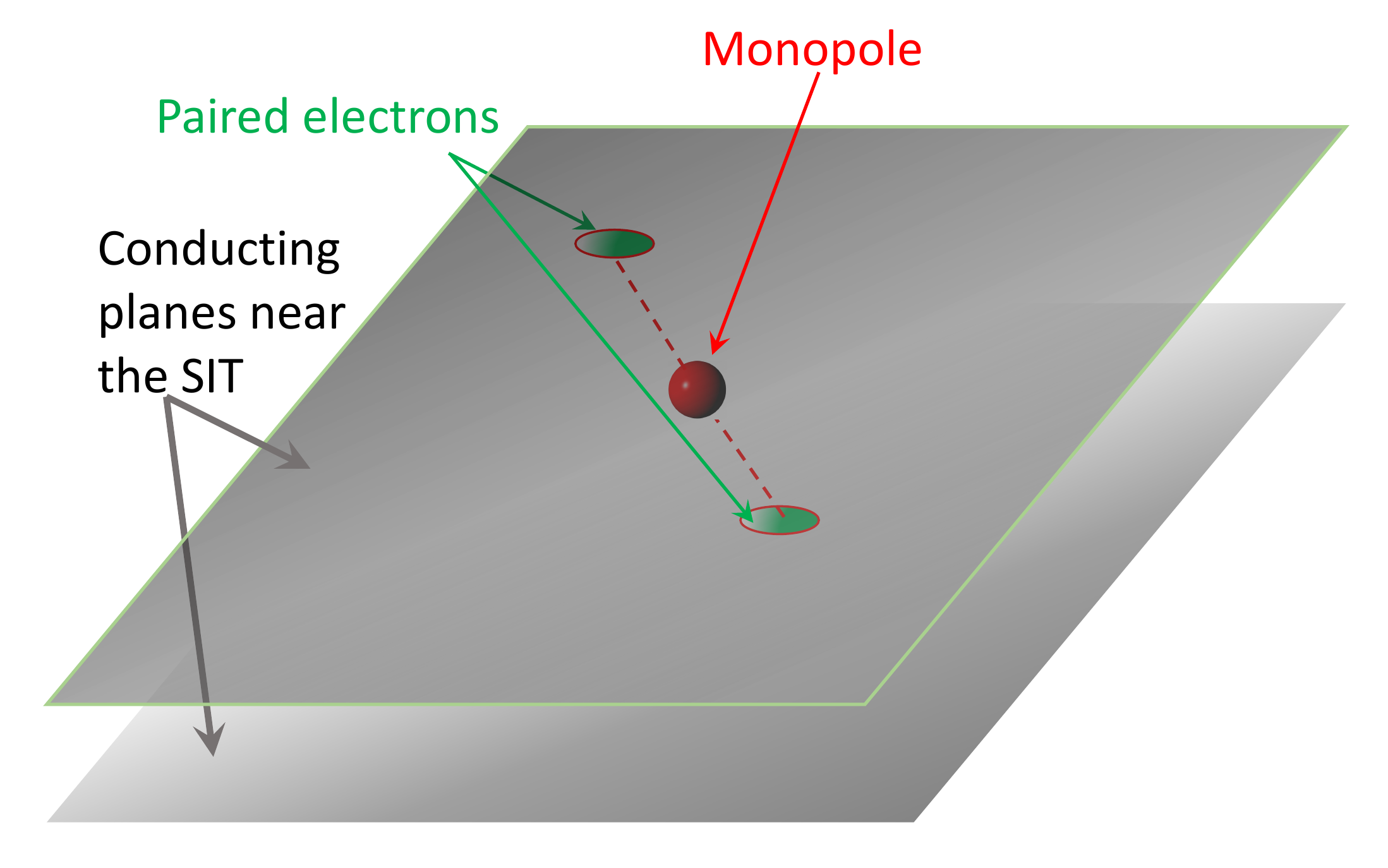}
	\vspace{-0.3cm}
	\caption{\textbf{A sketch of the minimal model for HTS pairing.} The conducting planes are shown in gray. Heavy monopoles appear in the middle between the planes and paired electrons are located on opposite planes, their motion being restricted to their respective planes.}
	\label{Fig1}
\end{figure}

Here we propose a new pairing mechanism straightforwardly leading to localized, strong coupling electron pairs. We propose pairing by effective magnetic monopoles forming between the conducting planes of a layered material. Magnetic monopoles bind two electrons residing in two respective adjacent conducting planes, see Fig.\,\ref{Fig1}, localizing the seed pairs around them. The strong coupling of the pair potential emerges from the topological Dirac quantization condition \cite{olive, preskill} which guarantees that the product of magnetic and electric charge is a coupling constant of order unity. This leads immediately to high critical temperatures.  The pairing model that we propose establishes the local character of pairs in the real space and predicts precisely that the pair size is smaller than their separation, explaining thus their emergence well above $T_c$, in accord with the experimental findings\,\cite{Bozovic2020}.

  \smallskip
  
Experiments\,\cite{Zhao2019} provide strong evidence for the proposed mechanism which pairs electrons locally and involves two conducting planes separated by atomic scales. Further experimental support  follows from the data of\,\cite{PNAS}, which have reported a rise of $T_{\mathrm c}$ in a cuprate superconductor when disorder eliminates the effect of charge density waves that suppress coupling between the adjacent CuO planes.

\section{2. Monopoles and gauge fields in condensed matter systems}
\noindent
It has been known for a long time that defects in condensed matter systems can be described by effective gauge fields\,\cite{kleinert}. 
Examples include curvature defects in graphene sheets\,\cite{pseudo1, pseudo2, pseudoplus1, pseudoplus2}, see\,\cite{pseudo3} for a review, and spin defects in magnetically ordered materials, like the cuprates\,\cite{rosch1, nagaosa, rosch2}.  In certain configurations, the defects appear to itinerant electrons like effective magnetic monopoles \cite{pseudo4, rosch2}. Here we show that such effective magnetic monopoles mediate a strong-coupling, real-space pairing mechanism which can explain hight-$T_c$ superconductivity. The pairing is localized around the effective monopoles and occurs well above the superconducting transition temperature $T_{\mathrm c}$.
\smallskip

In graphene sheets, strains, dislocations and curved protuberances, typically called ripplocations \cite{ripples1, ripples2, ripples3}, are equivalent to an effective gauge field coupled to the low-lying electronic degrees of freedom\,\cite{pseudo1, pseudo2, pseudoplus1, pseudoplus2} which typically amounts to a magnetic field $B= hK/4\pi e $ locally perpendicular to the sheet, with $K$ being the Gaussian curvature, see\,\cite{pseudo3} for a review. This field is referred to as pseudo magnetic field since it does not represent a real magnetic field but is an effective description of the geometric curvature effects induced by the defects. Most interestingly, it has been recently shown that the curvature of graphene nanobubbles is equivalent to a pseudo magnetic monopole at the center of the bubble\,\cite{pseudo4}. Note in this context that for general closed graphene sheets the Dirac quantization condition for the pseudo magnetic charge $g$ contained within the volume $V$ enclosed by the sheet $S$ is equivalent to the Gauss-Bonnet theorem, see e.g.,\,\cite{nash}, and, thus the magnetic charge is entirely fixed by the Euler number $\chi_{\rm E}$ (or the genus) of the closed sheet:
\begin{equation}
	g= g_{\rm Dirac}=  \int_V {\rm div}\  {\bf B} = \frac{h}{4\pi e} \int_S K = {h\over e} {\chi_{\rm E}\over 2} = {h\over e} (1-{\rm genus}) \ .
	\label{gauss}
\end{equation}

Here we consider configurations arising with two sheets, each one carrying a matching ``half-sphere" of a total defect, like ``two cupped hands". Note that such matching ``half-spheres" are the energetically favored configurations for realistic, strictly convex repulsive potentials among the sheets. Let us consider, thus, matching curvature defects of diameter $D$ in a system of two sheets separated by a distance $s$. In the extreme limit that $s\to 0$ outside the defect, this would correspond to a bubble with the topology of a sphere ${\cal S}^2$ and, according to (\ref{gauss}), the magnetic charge within this defect would be a Dirac monopole. There are corrections, however, due to the fact that the real defect is not a full sphere because of the finite inter-plane separation. Therefore, the defect can be continuously deformed to two planes by making the diameter smaller than the typical lattice spacing, or, in other words, the magnetic charge inside it can be extracted by moving it within the planes. To compute the magnetic charge of the defect we can first consider the total flux through a closed sphere, which is independent of the geometric details of the curvature, and then correct it by the magnetic flux flowing through a strip of radius $D/2$ and width $s$ around the equator, 
\begin{equation} 
	g= g_{\rm Dirac} \left( 1 -{s \over D} \right) \ .
	\label{magcharge}
\end{equation}
In the limit $s \to 0$ at fixed $D$ we recover the topological Dirac monopole. In the opposite limit $D \to s$ the magnetic charge vanishes because the defect has flattened out. In this paper we shall consider exclusively defects for which $s \ll D$ and $g\approx g_{\rm Dirac}$. Since there is no topological protection, it is an energetic question if such defects form or not. As we now show, they are indeed favoured energetically since they can lower the system's overall energy by forming Cooper pairs in their vicinity. To give a numerical estimate we use the value $g= 4.136 \times 10^{-15} \ {\rm Wb}$ for the charge of a single Dirac magnetic monopole to obtain the value $B= 3.3 \times 10^{4} \ T$ for the magnetic field on two planes separated by 1 {\AA}. 
\smallskip

Effective gauge fields appear also in magnetically ordered Mott insulators, like the CuO planes of cuprates. The motion of an itinerant electron in a non-collinear magnetic structure is subject to forces that can be represented as an effective compact $\mathrm{U(1)}$ gauge field\,\cite{rosch1, nagaosa, rosch2}. Quantized topological defects forming spin hedgehogs correspond then to emergent magnetic monopoles\,\cite{rosch2}. It is known that spin hedgehog defects with non-trivial topology can arise within Heisenberg antiferromagnets\,\cite{aoyama2021}. The Mott insulating state of the HTS cuprates is precisely such a Heisenberg antiferromagnet, and the idea is that spin-hedgehog defects arising in this state are seen by charges, that become increasingly itinerant upon doping, as effective magnetic monopoles. 
\smallskip

Common to all these effective Abelian monopole \cite{polyakov} configurations is that they require at least two planes in a layered material to be realized. In graphene sheets the two planes carry, by construction, the matching half-spheres forming an effective quasi-spherical bubble containing the quantized magnetic monopole (up to the corrections of $O(s/D)$). In magnetically ordered materials, the presence of effective magnetic monopoles also requires at least two planes, otherwise no hedgehog configuration would be possible. This condition is realized in cuprates, for example in BSCCO which is an exemplary HTS system, whose elementary cell contains, indeed, two CuO planes. 
The minimal model of a HTS material is thus taken as two conducting planes separated by the distance $s$ of the atomic scale. In material realizations, the two planes can be, e.g., the two CuO planes of cuprates or two graphene sheets in graphite. 
An effective magnetic monopole emerging between these planes binds two electrons, each residing at the respective opposite planes. The radical difference of the monopole-based binding mechanism from other commonly considered mechanisms\,\cite{Chubukov2005,Katsnelson2016,Chubukov2020-IV,Chubukov2020-3,HuangPairing,Rybicki} is that electron pairs are spatially localized around the monopoles. Heavy monopoles anchoring the electron pairs serve, thus, as nucleation points for a superconducting granular array that emerges upon cooling the system down from the temperature of pair formation, $T_{\mathrm{pair}}$ to $T_{\mathrm{c}}$. In a system supporting a sufficient monopole density, global superconductivity sets in when the droplets comprising the electron pairs and linked by tunnelling junctions form an infinite cluster. It thus occurs as the temperature compares with the coupling energy, and our estimate gives  $T_{\mathrm{c}} = {\cal O}\left( 10^2\right)$\,K for a typical granule size of ${\cal O}(1)$\,nm.  
We predict thus possible room-temperature superconductivity in layered materials in which the density of effective magnetic monopoles is sufficiently high. An appealing candidate is graphite, in which local superconductivity, concentrated around defects, has indeed been detected with critical temperatures of up to 300$^\circ {\rm K}$\,\cite{kopelevich1,kopelevich2,kopelevich3} and has been shown to form exactly the Josephson-junction-array-like structures\,\cite{Ballestar1,Ballestar2} that can lead to global superconductivity once the mechanisms establishing the global superconducting phase coherence set in.

\section{3. Electron pairing by a single monopole}
\noindent
In this section we present the detailed derivation of  pairing of two electrons confined to respective adjacent conducting planes by the magnetic monopole confined between these planes. This new pairing mechanism is the main result of our work.
\smallskip

Although they do not stem from real electromagnetism, the effective gauge fields induced by phenomena like curvature of conducting sheets or Berry phases in non-collinear magnetically ordered structures, couple to matter in the precisely same way as real electromagnetic gauge fields do. Hence, we can use a standard formalism like, e.g., Schr\"odinger equation in an external magnetic field, to derive the quantum effects of the action of the effective on electrons.
\smallskip

We start our derivation with considering two electrons interacting via the spherically symmetric repulsive $1/r$ Coulomb potential (we use natural units $c=1$, $\hbar = 1$, $\varepsilon_0 = 1$) and a short-range repulsion potential $V_{\rs R}(r)$. Decomposing the wave function into spherical harmonics yields the additional $\ell (\ell+1)/r^2$ repulsive centrifugal barrier for states with angular momentum $\ell >0$. If the short-range repulsion $V_{\rs R}(r)$ is stronger than the centrifugal barrier, the electrons settle into a state with sufficiently high angular momentum. Due to rotational symmetry, the $z$-component $m$, $|m| \le \ell$, of the angular momentum falls out of the Hamiltonian. What happens, however, if the spherical-symmetry-breaking mechanism encoded in a vector potential makes the Hamiltonian explicitly dependent on the $z$-component of angular momentum?  On dimensional grounds, this brings in an additional $1/r^2$ term that can have either negative or positive sign. If the sign is negative, it can counterbalance the centrifugal barrier resulting in an attractive $1/r^2$ potential at intermediate distances before the Coulomb repulsion takes over. Then the potential well forms, resulting in a discrete spectrum of the bound states with the finite angular momentum. 
\smallskip

Magnetic monopoles provide precisely such a spherical-symmetry-breaking mechanism\,\cite{olive}. In the presence of the magnetic monopole of the strength $g$, an electron acquires an additional angular momentum $L_{\rm M} = (eg/4\pi) \hat {\bf r}$, with $\hat r$ being the unit vector pointing from the monopole to the electron. The Dirac quantization, $eg=2\pi n$, $n\in {\mathbb Z}$, requires for this additional angular momentum contribution, originating from the interplay of the electric and magnetic fields of two point particles, to match the spectrum imposed by the rotation group. This is just a spherical symmetry breaking contribution to the angular momentum since it singles out the vector connecting the monopole to the electron. As we now show, this monopole-induced angular momentum can bind electrons, and the optimal angular momentum of the resulting pair depends on the monopole density. Lower densities favor higher angular momenta and vice versa. 
\smallskip

Importantly, monopoles change the statistics of original electrons, which become bosons themselves for odd $eg/2\pi$ but remain fermions for the even values of $eg/2\pi$\,\cite{preskill}. The exact centrifugal barrier cancellation takes place only for odd values of $eg/2\pi$. The overall centrifugal potential, however, vanishes or turns negative for all values of the total angular moment $\ell $ satifying $2\ell  \le | eg/2\pi |$. For such values magnetic monopoles induce pairing of electrons. Paired electrons can then Bose condense into droplets localized  near the monopole. Because of the Dirac quantization, magnetic monopoles are heavy excitations, with the mass $m_{\rs M}\propto 1/\alpha$, where $\alpha=e^2/\hbar c\approx 1/137$ is the fine structure constant. The droplets are anchored in space and mediate a pairing mechanism giving rise to superconductivity at elevated temperatures. Since the product of the electric and the magnetic charge is O(1) due to the Dirac quantization condition, this automatically provides the strong-coupling pairing mechanism without any further assumption. 

Note the dimensional dichotomy of these emerging high-temperature superconductors (EHTS) which they share wiht the cuprates. From the viewpoint of the charge- and magnetotransport, the EHTS materials exhibit a profoundly 2D behavior\,\cite{Pomar1996, Zhao2019}. However, phenomena related to topological aspects of the electronic spectrum, like the magnetoelectric effect in the pseudogap state\,\cite{kerr}, require the full underlying 3D microscopic nature. The proposed pairing mechanism is aligned with this dichotomy. The monopole pairing rests on the 3D quantum mechanical structure comprising two conducting planes separated by an atomic scale distance so that the charges can tunnel between the planes, providing robust Josephson links between them. At the same time, as long as the thermal coherence length $L_{\rs T}$$=$$\sqrt{2\pi D/(k_{\rs B}T)}$ exceeds the interplane distance $s$, the system exhibits two-dimensional transport properties. Here $D$$=$$(\pi/2\gamma)(k_{\rs B}T_{\mathrm c}/eB_{\mathrm c2}(0))$ is the electron diffusion length and $\gamma=1.781$ is Euler's constant. Since the dephasing length $L_{\phi}$$=$$\sqrt{D\tau_{\phi}}\gg L_{\rs T}$ as long as $k_{\rs B}T\gg \hbar/\tau_{\phi}$, the quasiparticle description holds well in this 2D electric response regime and the 3D quantum mechanical consideration of electron binding applies. This explains why a cuprate monolayer consisting of two conducting planes retains the same high transition temperature\,\cite{monolayer} as a 3D sample, whereas 2D films of conventional superconductors have $T_{\mathrm c}$ much lower than the bulk of the same material. 
\smallskip

Let us consider a heavy magnetic monopole of charge $g$ formed in the middle between the two conducting planes, see Fig.\,1. Using the remarkable result of\,\cite{Varlamov1993} that the quasiparticle lifetime within the layer is proportional to the intraplane scattering rate, we conclude that electrons are bound to intraplane motion with rare interplane hops, and that this intraplane electron motion becomes even more pronounced with increasing disorder and doping. We obtain a three-body quantum mechanical problem involving two electrons of charge $e$ restricted to respective parallel planes and showing the short-range repulsion, and the Dirac magnetic monopole of magnetic charge $g$ in between. We proceed with the simpler formulation of an infinitely heavy magnetic monopole located at the center of mass of the two-electron system. This reduces to a single-body problem of an electron of reduced mass $m/2$ in the external field of the magnetic monopole, which is amenable to an analytical, albeit approximate solution, maintaining the generality of the monopole pairing mechanism.
\smallskip

Dirac monopoles are particles carrying an attached Dirac string. If the string is aligned with the negative $z$-axis, the vector potential ${\bf A}_u$ of the monopole is\,\cite{olive} 
\begin{eqnarray}
	{\bf A}_u = f_u (r, \theta) \ \hat \varphi \ ,
	\nonumber \\
	f_u (r, \theta) = {g\over 4\pi r}  {1-{\rm cos}(\theta)\over {\rm sin} (\theta)} \ ,
	\label{gaugeupper}
\end{eqnarray}
which has a singularity at $\theta = \pi $. Here $r$, $\theta$ and $\varphi$ denote spherical coordinates and $\hat \varphi$ is the unit vector in the $\varphi$ direction. The value of the coupling constant $g$ will be taken henceforth as that of a fundamental magnetic monopole,  $g=4.136 \times 10^{-15} \ {\rm Wb}$. To solve the eigenstate problem in the field of a Dirac monopole one cannot use a single set of coordinates for the whole sphere but must use the Wu-Yang formalism\,\cite{wuyang} to cover the sphere with the so-called atlas of maps, supplemented by gauge transformation conditions on the overlap regions between the different maps. The simplest atlas comprises two maps, the upper hemisphere, $0 \le \theta \le \pi/2 + \epsilon$, with the gauge potential (\ref{gaugeupper}) and the lower hemishpere, $\pi/2 -\epsilon \le \theta \le \pi$, with the gauge-transformed potential
\begin{eqnarray}
	{\bf A}_l = f_l (r, \theta) \ \hat \varphi \ ,
	\nonumber \\
	f_l (r, \theta) = -{g\over 4\pi r}  {1+{\rm cos}(\theta)\over {\rm sin} (\theta)} \ ,
	\label{gaugelower}
\end{eqnarray}
corresponding to the same magnetic monopole at the origin of coordinates but with the Dirac string now along the positive $z$-axis
\begin{equation}
	{\bf A}_l = {\bf A}_u - \nabla \left( {g\over 2\pi} \varphi \right) \ .
	\label{gaugetr}
\end{equation}
In both hemispheres, the gauge potential is now regular allowing for solving corresponding Pauli equations. The price to pay is the gauge transformation connecting wave functions in the overlap region
$[\pi/2$$ -$$\epsilon, \pi/2$$+$$\epsilon] $. 
\smallskip

The Hamiltonian for two electrons with charges $e$ and masses $m$ and a fixed magnetic monopole with the magnetic charge $g$ at the origin is
\begin{eqnarray}
	H = {1\over 2m} \left( {\bf p}_1 - e{\bf A} ({\bf x}_1) \right)^2 + {1\over 2m} \left( {\bf p}_2 - e{\bf A} ({\bf x}_2) \right) ^2- {e\over m}{\bf s}_1\cdot {\bf B} \left( {\bf x}_1\right) 
	\nonumber \\
	- {e\over m} {\bf s}_2\cdot {\bf B} \left( {\bf x}_2\right)
	+V_{\rs R} \left( |{\bf x}_1 -{\bf x}_2 | \right) + V_{\rs C} \left( |{\bf x}_1 -{\bf x}_2 | \right) \ ,
	\label{ham}
\end{eqnarray} 
where ${\bf A} $ is the monopole gauge potential defined by Eqs.\,(\ref{gaugeupper}) and (\ref{gaugelower}), ${\bf s}_{1,2}$ denote the spin vectors of two electrons, $V_{\rs C}(r) = e^2/4\pi \varepsilon r$ is the repulsive Coulomb potential, with $\varepsilon$ the relative dielectric permittivity of the material and $V_{\rs R}(r)$ is the short-range repulsion. We introduce the center of mass and relative coordinates as ${\bf R} = ({\bf x}_1 + {\bf x}_2) /2$ and ${\bf r} = ({\bf x}_1 -{\bf x}_2)$ and we set ${\bf R} = 0$. To make the model amenable to an analytical solution we make further simplifying assumption that the infinitely heavy, external magnetic monopole sits exactly at the center of mass of the two-electron system. The time-independent Pauli equation becomes then
\begin{eqnarray}
	\biggl[ -{1\over 2m} \left( \nabla -ie{\bf A} \left( {{\bf r}\over 2} \right) \right)^2 -{1\over 2m} \left( \nabla +ie{\bf A} \left( {-{\bf r}\over 2} \right) \right)^2 -{|eg|\over 2\pi m |{\bf r}|^2} 
	\nonumber \\
	+V_{\rs R} (|{\bf r}|) + V_{\rs C} (|{\bf r}|) \biggr] \psi = E \psi \ ,
	\label{sch1}
\end{eqnarray}
where we have specialized to a total spin 0 state in which the spin of each electron has a hedgehog configuration parallel or antiparallel to the monopole magnetic field, depending on the sign of $g$. 
\smallskip

Suppose now that the motion of electrons is constrained to the two conducting planes at $z$$=$$\pm s$, with total angular momenta $\pm \ell$, $2\ell \in {\mathbb N}$, on the upper and lower planes, respectively. If the monopole charge satisfies $|eg/2\pi | = 2\ell$ condition, Eq.\,(\ref{sch1}) reduces to a single 2D radial equation, see Appendix,
\begin{equation} 
	\biggl[ -{1\over m} \left(  {1\over x} {\partial \over \partial_x}  \left( x {\partial \over \partial x} \right) \right) + V_{\rs R}(x) 
	-{|eg/2\pi | \over m(s^2 + x^2)}  +V_{\rs C}(x) \biggr] F(x) = E F(x) \ ,
	\label{unieq}
\end{equation}
where $x= r\ {\rm sin}(\theta)$ is the radial distance on the planes and $V_{\rs C}(x) = {\alpha} / \varepsilon \sqrt{s^2 + x^2}$. 
The short-range repulsion models, e.g., the quantum statistical pressure of electrons when they are squeezed by two conducting planes. Its exact form does not matter; however, it forces electrons to fall into the non-zero orbital angular momentum states $\ell  > 0$ in order to avoid the energy price to be too close: $F(x) \propto x^\ell $ for $x\ll s$ and the higher $\ell $, the more the wave function is suppressed at the origin. Yet, for $|eg/2\pi / = 2\ell$, the resulting repulsive centrifugal barrier gets completely canceled by the additional, monopole-induced angular momentum, and only an attractive interaction due to the electron magnetic moments survives. If $|eg/2\pi | > 2\ell$, which can occur only for higher magnetic charges, the centrifugal ``barrier" turns negative and the attraction gets even stronger. 
So, for $|eg/2\pi | \ge 2\ell$, the potential well forms between the two relevant scales $s$ and $ a= \varepsilon /m\alpha$, where the Coulomb repulsion takes over, and the electrons form pairs that can Bose condense in a droplet localized around the positions of the heavy monopole. To estimate the optimal value of $\ell$, note that, on one hand, increasing $\ell$ makes the attraction stronger, and on the other hand, the higher the monopole charge, the heavier they are and the higher the energy cost of creating them between the planes. Given the large mass of monopoles, one expects that at small monopole densities larger values of $\ell$ are favored and vice versa. In any case, the proposed mechanism accommodates all types and the possibilities of pairing,  including s-wave, p-wave and d-wave pairings, depending on the magnetic charge of the pseudo-magnetic monopoles forming between the planes. 
\smallskip

The construction of the potential well results again from the interplay of dimensionalities. The angular momentum of the electrons constrained to two planes is a 2D effect. The additional angular momentum due to the monopole and the magnetic moment interactions, however, are the 3D effects, since they are directed from the monopole at the center to the locations of the electrons on the planes. At sufficiently large distances, the additional angular momentum cancels out the centrifugal barrier and the magnetic moment interaction causes the overall attraction. 
\smallskip

The potential well is determined by the combination of the magnetic moment attraction with the short-range repulsion representing quantum statistical pressure of electrons of inter-layer atoms squeezed between the two conducting planes. As such, it should be a scale-free, i.e., $1/x$ potential. Therefore, in general, both the position of the minimum of the potential well and the bound state energy, are functions of the two spatial scales $s$ and $a$. If they are comparable, there remains only a single spatial scale and, accordingly, a single energy scale. To estimate it, one can neglect $V_{\rs R}(x)$ and $V_{\rs C}(x)$ in Eq.\,(\ref{unieq}). By multiplying the whole equation by $m$ we see that the scale $mE$ in the right-hand side is determined by the unique remaining scale $s$ in the left hand-side. Therefore $E_0 = {\cal O}(1/ms^2)$. As shown in\,\cite{VortexReview}, the numerical coefficient is of order one so that 
\begin{equation}
	E_0 \simeq\frac{\hbar^2}{ms^2}\, ,
	\label{gap}
\end{equation}
where we restore physical units. 
Since the interplane spacing $s$$\simeq$$k_{\rs F}^{-1}$, with $k_{\rs F}$ being the Fermi wavevector, $E_0$$=$${\cal O}(E_{\rs F})$. 
The corresponding localization size of the bound state within the plane is $\ell_{\parallel}\simeq s$, and as discussed above it is the same orthogonal to the plane, $\ell_{\perp}$$\simeq$$s$. 
\smallskip

We have until now discussed only the electron pairing by a single monopole. A complete model, however, should also take into account how these potential wells behave in presence of many monopoles. A full many-body theory lies beyond the scope of this paper but one give qualitative arguments for the changes introduced by many monopoles. At finite monopole density $\rho$, when monopoles are brought near each other, the binding energy $E_0$ first keeps increasing as long as the
two-dimensional, parallel to the plane inter-monopole distance $d=(\rho s)^{-1/2}>\ell_{\parallel}\approx s$\,\cite{Baz'}. This is the effect of the 2D Lifshitz localization\,\cite{Nelson1993}. However, at $d<s$, the ``flat" bottoms of the potential wells in the many-monopole generalization of Eq.\,(\ref{unieq}) overlap. Then the pairs are localized by fluctuations in the monopole density rather than by single potential wells, see\,\cite{Nelson1993}, where a similar problem was discussed, and $E_0$ starts to decrease with increasing $\rho$. Therefore, one expects that the optimal binding is achieved at $d\simeq s$. This prediction for the optimal monopole density is in accord with the recent experimental data of\,\cite{pressure} showing that the highest transition temperature in carbonaceous sulfur is indeed achieved at some optimal distance $s$ between the conducting planes. An exact calculation of $E_0$ would require a microscopic treatment, but for present purposes the estimate (\ref{gap}) is sufficient.
\smallskip

Here, a comment is in order. Our proposed pairing mechanism is predicated upon the strong residual attractive Pauli interaction induced by the field of the pseudo-magnetic monopole on the two-electrons state with the total spin $S_{\rm tot} = 0$ after the same pseudo-monopole field cancels the centrifugal barrier. The resulting Maki parameter, describing the relative importance of the orbital- and Pauli pair breaking mechanisms, $\alpha_{\mathrm M}\simeq E_0/E_{\mathrm F}$, is of order unity. One may expect then that the magnetic field-induced depairing occurs at fields exceeding the Pauli limit for conventional superconductors and that the re-entrant superconductivity similar to that observed in trilayer graphene\,\cite{Pablo2021} and, possibly, in\,\cite{Barzola}, may develop as a result of the monopole pairing. The effects of high magnetic fields, however, require a detailed investigation and will be the subject of the forthcoming publication.

\section{4. Discussion and conclusion}

\noindent
The localized pairs with the binding energy $E_0$ are the nucleation centers for superconducting droplets. As we have derived above, the $T=0$ dimension of such droplets is $s$. At higher temperature the droplet dimension $\xi (T)$ will typically be larger. 
Global superconductivity sets in at the temperature when sufficient monopoles have formed so that these droplets form an infinite three-dimensional cluster and is given by the 
Ioffe-Larkin\,\cite{Ioffe1981} formula for the transition temperature $T_{\mathrm c}$ in highly inhomogeneous superconductors
\begin{equation} 
	k_{\rs B} T_{\mathrm c} = \omega\  {\rm e}^{-0.89 {d/\xi(T_{\mathrm c} )}} \ ,
	\label{lif}
\end{equation}
where $\xi(T)$ is the characteristic size of the localized pair and $\omega$ is the attempt frequency in the matrix element $t=\omega\exp(-d/\xi)$ describing tunneling of the bound pair between adjacent monopoles. The mean inter-monopole distance is itself a function of temperature and material characteristics, but, for an estimate, we take the optimal monopole density providing the strongest binding, i.e. $d\simeq s$. To favor the factors that can lower the expected $T_{\mathrm c}$, however, we take the smallest possible size of the pair, i.e.  $\xi \simeq s$. The tunneling matrix element between monopoles at the distance $d$ is found following\,\cite{Nelson1993}, and gives $\omega=\sqrt{8/\pi}(\hbar^2/ms^2)\sqrt{s/d}$.
The resulting estimate for the transition temperature is 
\begin{equation}
	T_{\mathrm c} \approx 0.65\cdot[\hbar^2/({ms^2k_{\rs B}})]\,,
	\label{est}
\end{equation}
where we have restored physical units. Taking the interplane distance $s$ as\,1\,nm, one obtains  $T_{\mathrm c} $$=$$ {\cal O}(10^2)\,K$. 
\smallskip

The proposed real-space monopole pairing mechanism reveals the microscopic nature of HTS. It solves the puzzle why the pairing size in HTS does not exceed the inter-pair distance; this follows from the fact that, since the optimal $d\simeq s$, then, in general, $d\gtrsim s$. Therefore, the distance between paired electrons is larger than the pair size in HTS. Our findings pave the way for tailoring superconducting materials with enhanced $T_{\mathrm c}$. To that end, one has to maximize the number of curvature defects on the adjacent conducting planes while minimizing the inter-plane distance $s$. The elevated curvature density promotes an enhanced generation of monopoles, thereby lowering the system's overall energy. Determining the optimal monopole density and finding the corresponding optimal material parameters requires a detailed self-consistent microscopic treatment of the intertwined electron-monopole ensemble and the derivation of the superconducting transition temperature $T_{\mathrm c}$. This self-consistent microscopic theory will be the subject of a forthcoming publication. Here we conclude that our estimate predicts the possibility of realizing room-temperature superconductivity in layered compounds like graphite and cuprates or similar with the sufficient density of dopants or topological defects providing the sufficient density of topological curvature centers hosting monopoles.

\smallskip
\textit{Acknowledgments}–
We are delighted to thank Andrey Chubukov, Alexander Golubov, and Y. Kopelevich for illuminating discussions. The work by V.M.V. was supported by Terra Quantum AG.
M.C.D. thanks CERN, where she completed this work, for kind hospitality.


\section*{Appendix: Derivation of the Pauli equation on the planes}
\noindent
Using that the monopole gauge potentials (\ref{gaugeupper}) or (\ref{gaugelower}) are divergenceless, we can simplify equation (\ref{sch1}) of the main text to
\begin{eqnarray}
	&&\biggl[ -{1\over m} \nabla^2 +{ie\over 2m} \left( {\bf A} \left( {{\bf r}\over 2} \right) \cdot \nabla - {\bf A} \left( {-{\bf r}\over 2} \right) \cdot \nabla \right)+ {e^2\over 2m} \left( {\bf A}^2 \left( {{\bf r}\over 2} \right)+ {\bf A}^2 \left( {-{\bf r}\over 2} \right) \right)
	\nonumber \\
	&& -{|eg|\over 2\pi m |{\bf r}|^2}  +V_{\rs R} (|{\bf r}|) + 
	V_{\rs C}(|{\bf r}|) \biggr] \psi = E\psi \ .
	\label{sch2}
\end{eqnarray}
Starting from this generic equation, one has to formulate two Pauli equations, one for each hemisphere, as explained in the main text. Let us begin with the upper hemisphere, denoted by the subscript ``$u$". Since we restrict to values $0\le \theta \le \pi/2+ \epsilon$, the arguments of the second gauge potentials ${\bf A}$ in (\ref{sch2}) relate to the lower hemisphere, denoted by subscripts ``$l$". Therefore, we have to use Eq.\,(\ref{gaugeupper}) for the first instance of the gauge potential and Eq.\,(\ref{gaugelower}) for the second one. This gives
\begin{eqnarray}
	&&\biggl[ -{1\over m} \nabla^2 +{ie\over 2m} \left( f_u \left( {r\over 2}, \theta \right) - f_l \left( {r\over 2}, \pi-\theta \right) \right) 
	{1\over r{\rm sin}(\theta)} {\partial \over \partial \varphi} 
	\nonumber \\
	&&+ {e^2\over 2m} \left( f^2_u \left( {r\over 2}, \theta \right) + f^2_l \left( {r\over 2}, \pi-\theta \right) \right) -{|eg|\over 2\pi m r^2}\nonumber \\ &&+V_{\rs R}(r ) +V_{\rs C}(r) \biggr]  \psi_u = E\psi_u \ .
	\label{sch3}
\end{eqnarray}
Repeating the same reasoning for the lower hemisphere $\pi/2 -\epsilon \le \theta \le \pi$, we obtain the second Pauli equation
\begin{eqnarray}
	&&\biggl[ -{1\over m} \nabla^2 +{ie\over 2m} \left( f_l \left( {r\over 2}, \theta \right) - f_u \left( {r\over 2}, \pi-\theta \right) \right) 
	{1\over r{\rm sin}(\theta)} {\partial \over \partial \varphi}
	\nonumber \\
	&&+ {e^2\over 2m} \left( f^2_l \left( {r\over 2}, \theta \right) + f^2_u \left( {r\over 2}, \pi-\theta \right) \right) -{|eg|\over 2 \pi m r^2}\nonumber \\ &&+ V_{\rs R}(r) + V_{\rs C}(r) \biggr]  \psi_l = E\psi_l \ .
	\label{sch4}
\end{eqnarray}
Using (\ref{gaugeupper}) and (\ref{gaugelower}), we obtain, finally, the explicit expressions of our pair of Pauli equations
\begin{eqnarray}
	\biggl[ -{1\over m} \nabla^2+i {(eg/2\pi)\over m r^2} 
	{1-{\rm cos}(\theta)\over {\rm sin}^2(\theta)} {\partial \over \partial \varphi} 
	+ {(eg/2\pi)^2\over mr^2} \left( {1-{\rm cos}(\theta) \over {\rm sin}(\theta)} \right)^2 
	\nonumber \\
	-{|eg|\over 2 \pi m r^2} + V_{\rs R}(r) + V_{\rs C}(r) \biggr]  \psi_u = E\psi_u \,,
	\nonumber \\
	\biggl[ -{1\over m} \nabla^2-i {(eg/2\pi)\over m r^2} 
	{1+{\rm cos}(\theta)\over {\rm sin}^2(\theta)} {\partial \over \partial \varphi} 
	+ {(eg/2\pi)^2\over mr^2} \left( {1+{\rm cos}(\theta) \over {\rm sin}(\theta)} \right)^2 
	\nonumber \\
	-{|eg|\over 2 \pi m r^2} + V_{\rs R}R(r) + V_{\rs C}(r) \biggr]  \psi_l = E\psi_l \ .
	\label{sch5}
\end{eqnarray}
The presence of the magnetic monopole is reflected in three new terms, the first two of which, as anticipated, break the spherical symmetry of the original Coulomb problem. The first embodies the monopole-induced additional contribution to the z-axis component of the angular momentum: it has a different sign in the upper and lower hemispheres since, as we discussed above, it points from the monopole to the electrons and the monopole sits exactly in the middle. Its coefficient is only weakly dependent on $\theta$ since it varies from 1 on the equator to 1/2 at the poles. The second new term, instead is a repulsive term concentrated around the equator and vanishing near the poles. Finally, the third new term is the magnetic attraction due to electron magnetic moments. 

Since the vector gauge potentials in the lower and upper hemispheres are gauge transforms of each other, we must impose the Wu-Yang gauge conditions also on the wave functions in the overlap region $[\pi/2 -\epsilon, \pi/2 + \epsilon] $ of the maps of the atlas\,\cite{wuyang}, 
\begin{equation}
	\psi_l = {\rm e}^{-i{eg\over 2\pi}\varphi} \psi_u \ .
	\label{wuyang}
\end{equation}
We can thus make the Ansatz 
\begin{eqnarray}
	\psi_u \left( r, \theta, \varphi \right)= {\rm e}^{+i {eg\over 4\pi} \varphi} F_u (r, \theta, \varphi )  \ ,
	\nonumber \\
	\psi_l \left( r, \theta, \varphi \right)= {\rm e}^{-i {eg\over 4\pi} \varphi} F_l (r, \theta, \varphi )  \ .
	\label{ansatz1}
\end{eqnarray}
Because an exchange of the two electrons involves necessarily also a swap of hemispheres, the exchange operator on the wave function must take into account the Wu-Yang gauge transformation\,\cite{wuyang}. Therefore, for $eg/2\pi$ an odd integer the exchange implies a factor (-1) and the statistics of the electrons is changed to bosons. For $eg/2\pi$ an even integer, instead there is no additional (-1) factor and the statistics of the individual electrons remains fermionic. Correspondingly, for $eg/2\pi$ an odd integer the additional gauge factor is a double covering representation of $2\pi$ rotations, while it is single-valued for $eg/2\pi$ an even integer. This is the statistical transmutation induced by magnetic monopoles (for a review see \cite{olive}) . Independently of this statistical transmutation of the individual components, however, the total spin 0 pair is a boson. 

We now constrain the electron motion to two parallel horizontal planes at $z=\pm s$ with the monopole at the origin. As a consequence, the Laplace operator reduces to
\begin{equation}
	\nabla^2 = {\partial^2 \over \partial x^2} + {1\over x} {\partial \over \partial x} +{1\over x^2} {\partial^2 \over \partial \varphi^2} \ ,
	\label{laplaceplane}
\end{equation}
where $x$ denotes the radial distance on the two planes. We are interested primarily in small values of $x$ and the second new term in (\ref{sch5}), ${\cal O}(\theta^2)$, is subdominant with respect to the first one, ${\cal O}(1)$ near the poles: we will henceforth neglect it. In addition, since the electrons are forced to move on the two horizontal planes, we do not have to use the usual monopole harmonics\,\cite{wuyang} but we can make use of the much simpler cylindrical harmonics decomposition by making the Anstaz 
\begin{eqnarray}
	F_{u} (r, \theta, \varphi)= {\rm e}^{-i \ell \varphi} F(x)  \ ,
	\nonumber \\
	F_{l} (r, \theta, \varphi )= {\rm e}^{+i \ell  \varphi} F(x)  \ ,
	\label{ansatz2}
\end{eqnarray}
where $r$ and $\theta$ are bound by the condition $x$$=$$ r{\rm sin} \theta$ and $\ell$, $2\ell \in {\mathbb N}$ is the total angular momentum. The $z$-components of the angular momentum of the electrons on the two planes cancel out, but the total angular momentum $\ell $ can well be different from zero. It is this value that indicates how much the axis of the composite wave function is tilted with respect to the $z$-axis. Combining (\ref{ansatz2}) with (\ref{ansatz1}) gives the announced result. When $| eg/2\pi |= 2\ell$ the effective angular momentum and the ensuing centrifugal barrier vanish altogether (note that for negative values of the magnetic charge the two equations (\ref{ansatz2}) are interchanged). We obtain thus a single radial equation for both planes, Eq.\,(\ref{unieq}) of the main text. 

\section*{Data availability}
Data sharing is not applicable to this article as no datasets were generated or analyzed during the current study.


\bigskip

\hskip 1pt

\end{document}